\documentclass[12pt, draftclsnofoot, onecolumn]{IEEEtran}	% regular single-column
 \usepackage{amsmath,amssymb}
 \usepackage{subfigure}
 \usepackage{graphicx,graphics,color,psfrag}
 \usepackage{cite,balance}
 \usepackage{algorithm}
 \usepackage{accents}
 \usepackage{amsthm}
 \usepackage{bm}
 \usepackage{url}
 \usepackage{algorithmic}
 \usepackage[english]{babel}
 \usepackage{multirow}
 \usepackage{enumerate}
 \usepackage{cases}
 \usepackage{stfloats}
 \usepackage{dsfont}
 \usepackage{color,soul}
 \usepackage{amsfonts}
 \usepackage{cite,graphicx,amsmath,amssymb}
 \usepackage{subfigure}
 \usepackage{fancyhdr}
 \usepackage{hhline}
 \usepackage{graphicx,graphics}
 \usepackage{array,color}
 \usepackage{amsmath}
 \usepackage{amsthm}
  \usepackage{booktabs}
\usepackage{framed}

\newtheorem{remark}{\bf Remark}
\def\phi{\varphi}

\def\l{\left}
\def\r{\right}
\def\({\left(}
\def\){\right)}

\def\bg{{\mathbf{g}}}

\def\bs{{\mathbf{s}}}

\def\bw{{\mathbf{w}}}

\def\by{{\mathbf{y}}}
\def\bz{{\mathbf{z}}}
\def\b0{{\mathbf{0}}}

\def\bG{{\mathbf{G}}}
\def\bH{{\mathbf{H}}}

\def\bL{{\mathbf{L}}}

\def\bU{{\mathbf{U}}}
\def\bV{{\mathbf{V}}}

\def\bX{{\mathbf{X}}}
\def\bY{{\mathbf{Y}}}

\begin{document}

\title{\huge Fast  Analog Transmission for High-Mobility Wireless Data Acquisition in Edge Learning}
\author{Yuqing Du and Kaibin Huang (The University of Hong Kong)

%\thanks{ Y. Du and K.~Huang are with  The University of Hong Kong, Hong Kong (Email: yqdu@eee.hku.hk, huangkb@eee.hku.hk).}
}

\maketitle

\vspace{-45pt}

\begin{abstract}
By implementing machine learning at the network edge, edge learning trains models by leveraging rich data distributed at edge devices (e.g., smartphones and sensors) and in return endow on them capabilities of seeing, listening, and reasoning. In edge learning, the need of high-mobility wireless data acquisition arises in scenarios where edge devices (or even servers) are mounted on ground or aerial vehicles. In this paper, we present a novel solution, called \emph{fast analog transmission} (FAT), for high-mobility data acquisition in edge-learning systems, which has several key features.   First, FAT incurs low-latency. Specifically, FAT requires no source-and-channel coding and no channel training  via   the   proposed technique of \emph{Grassmann analog encoding} (GAE) that encodes  data samples into subspace matrices. Second, FAT supports spatial multiplexing by directly transmitting analog vector data over an antenna array. Third, FAT can be seamlessly  integrated with edge learning (i.e., training of a classifier model in this work). In particular, by applying  a Grassmannian-classification algorithm from computer vision, the  received GAE encoded data can be  directly applied to training the model without decoding and conversion. This design  is found by simulation to outperform conventional schemes in learning accuracy  due to its robustness against  data distortion induced by fast  fading.

\end{abstract}

%By implementing machine learning at the network edge, edge learning trains models by leveraging rich data distributed at edge devices (e.g., smartphones and sensors) and in return endow on them capabilities of seeing, listening, and reasoning. In edge learning, the need of high-mobility wireless data acquisition arises in scenarios where edge devices (or even servers) are mounted on ground or aerial vehicles. In this paper, we present a novel solution, called \emph{fast analog transmission} (FAT), for high-mobility data acquisition in edge-learning systems, which has several key features. First, FAT incurs low-latency  via the proposed technique of \emph{Grassmann analog encoding} (GAE) that encodes  data samples into subspace matrices. Second, FAT supports spatial multiplexing. Third, FAT can be seamlessly  integrated with edge learning. In particular, by applying  a Grassmannian-classification algorithm, the received GAE encoded data can be directly applied to training the model without decoding and conversion. This design  is found by simulation to outperform conventional schemes in learning accuracy  due to its robustness against  data distortion induced by fast  fading. 

% \begin{figure}[t]
%  \centering
%\includegraphics[width=0.3\textwidth]{high_mobility_scenario_v4.pdf}   \vspace{-10mm}
%     \caption{A scenario of high-mobility wireless data acquisition for  edge learning where edge devices are mounted on ground vehicles or \emph{unmanned aerial vehicles} (UAVs).}
%
%  \label{Fig:High mobility scenario}
%\end{figure}

%\vspace{-8pt}
\vspace{-3mm}
\section{introduction}
\begin{figure}[tt]
  \centering
 \vspace{-1mm}\subfigure[]{\includegraphics[width=0.4\textwidth]{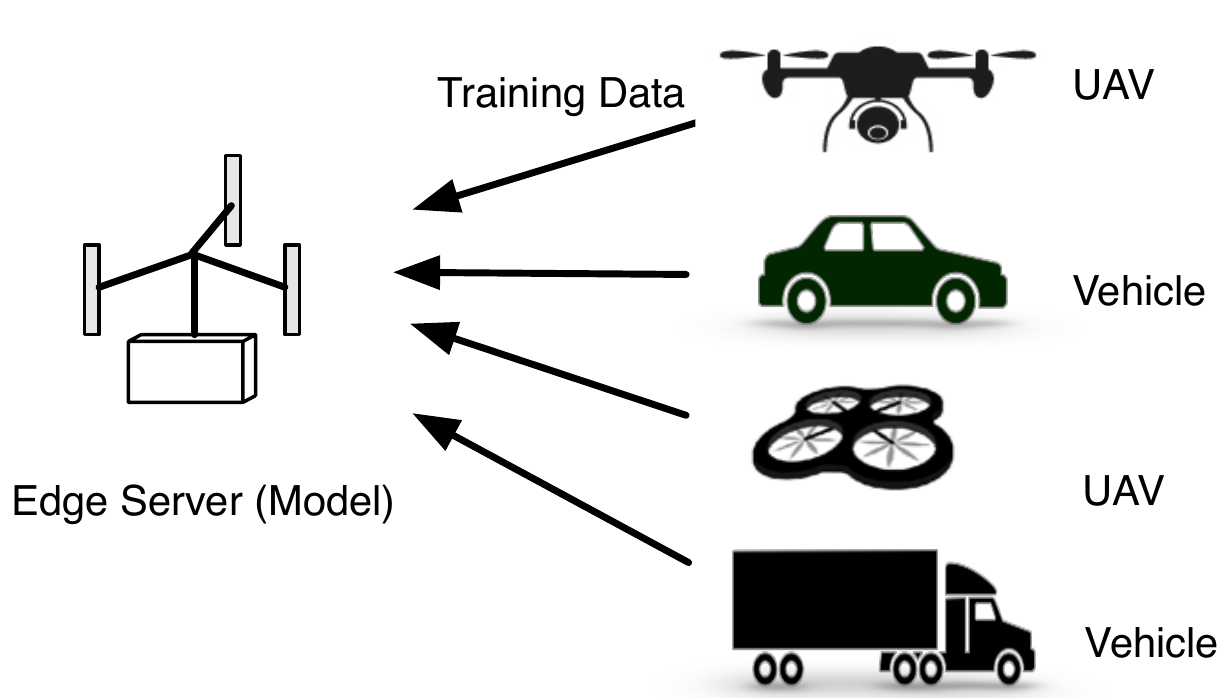} \label{Fig:High mobility scenario}}
 \vspace{-1mm}\subfigure[]{\includegraphics[width=0.55\textwidth]{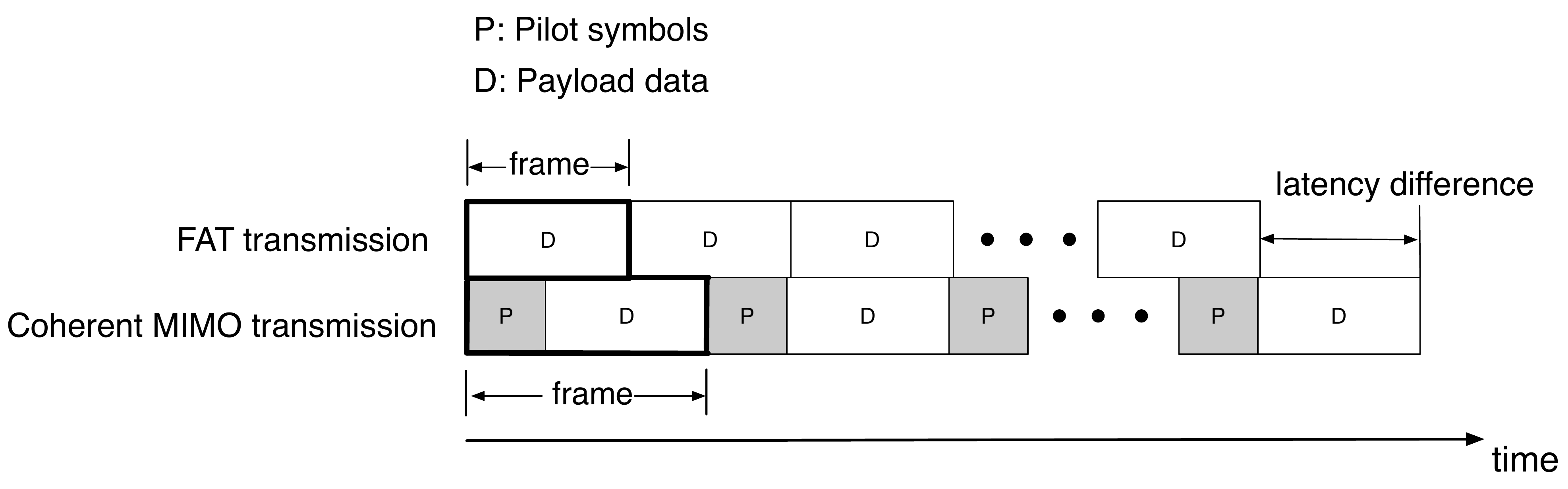}\label{Fig:latency} }
 %\vspace{-1mm}
     \caption{(a) A scenario of high-mobility wireless data acquisition for  edge learning where edge devices are mounted on ground vehicles or \emph{unmanned aerial vehicles} (UAVs); (b) Illustration of communication latency caused by channel training.}
 \vspace{-1mm}
  \label{Fig:datastructure}
\end{figure}

\vspace{-2mm}
Envisioned as an evolution in computing, \emph{edge learning} refers to the implementation of machine learning at the network edge so as to leverage enormous data distributed at edge devices (e.g., smartphones and sensors) for training models \cite{konevcny2015federated, mcmahan2016communication}.  Subsequently, the models are applied to empowering edge devices with the capabilities of seeing, listening and reasoning. While computing speeds are growing rapidly, the  latency in wireless data acquisition  has emerged  to be the bottleneck of fast edge learning~\cite{mcmahan2016communication}.  This issue is exacerbated in high-mobility scenarios where edge devices (or even edge servers)  are mounted on ground or aerial vehicles as illustrated Fig.~\ref{Fig:High mobility scenario} \cite{bockelmann2016massive}. High-mobility data acquisition faces several challenges: 1) robustness against fast fading, 2) low-latency given short connection time, 3) seamless integration with learning algorithms. To tackle these challenges, we present  a novel solution, called \emph{fast analog transmission} (FAT).

The design of the FAT scheme builds on three  ideas from  communication and  learning.
The first idea is \emph{analog transmission} based on linear analog modulation that has been deployed previously in different settings such as fast transfer of \emph{channel-state information} (CSI)~\cite{marzetta2006fast} and over-the-air functional computation  in sensor networks~\cite{goldenbaum2013robust}. Compared with digital transmission, the analog design does not require source-end-channel coding and decoding, thereby reducing computation  complexity.  Moreover, direct transmission of analog data instead of a quantized bit stream shortens the transmission duration. In terms of learning performance, our findings suggest that a customized design of analog transmission targeting learning (e.g., FAT) can be more robust than digital counterparts against data distortion by fast fading at high mobility.

The second idea is \emph{blind multiple-input-multiple-output (MIMO) transmission} without CSI. This idea was first developed in the classic area of non-coherent MIMO, which is a digital space-time modulation scheme \cite{hochwald2000unitary, yang2013capacity}. Its unique feature is a modulation constellation comprises a set of  subspace matrices. The  transmitted space-time symbol in the form of such a matrix is invariant to  rotation by a block fading  channel that  remains  constant within each symbol duration but varies over different durations. Thus the matrix can be transmitted and detected even without CSI at either  side, referred to hereafter as the \emph{channel-invariant property}  \cite{hochwald2000unitary}. Consequently, channel training is unnecessary, thereby reducing the transmission latency and overhead as illustrated in Fig.~\ref{Fig:latency}. On the other hand, non-coherent MIMO cannot support spatial multiplexing like its coherent counterpart. The resultant low data rates makes the former less popular in practice and its applications are limited to low-rate ultra-fast machine-type applications~\cite{bockelmann2016massive,boccardi2014five}. In contrast, the proposed scheme retains the advantages of both technologies, namely \emph{channel-invariant   property} and (analog) \emph{spatial multiplexing}. The said property  is achieved by the proposed  \emph{Grassmannian analog encoding} (GAE), a key FAT component,  which  encodes a data sample (an analog vector) into  a subspace  matrix by projection onto  a point on the Grassmann manifold,  thereby giving the name of the technique. On the other hand, FAT supports spatial multiplexing by directly transmitting analog data vectors instead of a single constellation point as in  non-coherent MIMO. 
The differences between  FAT and conventional MIMO schemes are summarized in Table~\ref{Table:Compare}. 
% \begin{figure}[t]
%  \centering
%\includegraphics[width=0.45\textwidth]{fig_latency.pdf}  \vspace{-5mm}
%     \caption{Illustration of communication latency caused by channel training.}
% 
%  \label{Fig:latency}
%\end{figure}

In this paper, we consider a typical edge-learning task of training a classifier model. The last idea  pertains to edge  learning and is to apply a Grassmann classification algorithm for classifying the received GAE encoded  training data. Such algorithms were originally developed for computer vision where image features or motions are represented as subspaces or equivalently points on a Grassmann manifold, referred to as \emph{Grassmann data}~\cite{hamm2008grassmann}. Via the application of such an algorithm, classification  can be seamlessly integrated with FAT since the  received GAE encoded data can be directly used in learning without decoding and conversion. Furthermore, the integration leads to accurate edge learning with robustness in data acquisition against fast fading.

 \begin{table}
\centering
\caption{Comparison of different MIMO transmission schemes}
\footnotesize
\begin{tabular}{cccccc}
\toprule
Transmission Scheme & Modulation & Signal Content & Blind Detection & Spatial Multiplexing &Latency\\
\midrule
FAT (proposed) & Analog & Data Coefficients & Yes & Yes &Ultra-Fast\\
Coherent MIMO & Digital & Bits & No & Yes & Slow\\
Non-coherent MIMO & Digital & Bits & Yes & No & Fast\\
\bottomrule
\end{tabular}
\label{Table:Compare}
\end{table}

 In summary, an edge learning system based on FAT comprises  the following {\bf three components} (elaborated in Section~\ref{MainSection}). 
\begin{enumerate}
\item \emph{Grassmann analog encoding:} At each edge  device, the proposed GAE encodes data samples into \emph{subspace} matrices by projection onto  a Grassmann manifold to enable blind MIMO transmission and robust edge learning.
\item \emph{Analog transmission and detection:} The GAE encoded data is transmitted using \emph{linear analog modulation} and blindly  detected at the edge server without channel knowledge. 
\item \emph{Edge learning:} At the edge server, the received Grassmann  data is used for training a classifier model  using a \emph{Grassmann-classification} algorithm from computer vision. 
\end{enumerate} 

By evaluating   the classification  performance of a model   and transmission latency using  simulation (see Section~\ref{Simulation}), the proposed FAT scheme  is found to substantially outperform the conventional  coherent (analog and digital) MIMO transmission at high mobility. 

 \vspace{-3mm}
\section{System and Simulation  Models}
\vspace{-2mm}

\subsection{System Model} 
\begin{figure}[t]
  \centering
\includegraphics[width=1\textwidth]{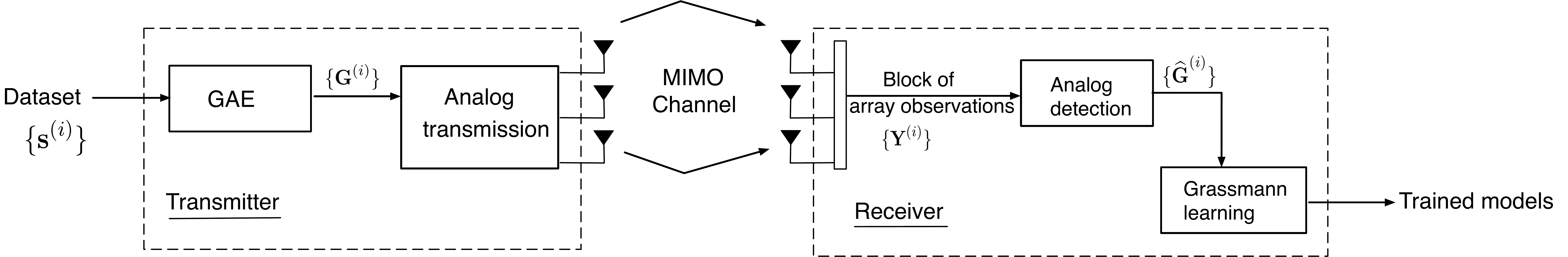} 
\caption{An edge-learning system based on  FAT.  }
 % \vspace{-1mm}
  \label{Fig:Grassmann_encoding_scheme}
\end{figure}

\vspace{-2mm}
Consider the edge-learning  system  illustrated in Fig.~\ref{Fig:Grassmann_encoding_scheme} where an edge server trains a classifier using  a training dataset transmitted by multiple edge devices. The transmissions by devices are based on time sharing and independent of channels given no CSI.  All nodes are equipped with antenna arrays, resulting in a set of narrow-band MIMO channels. Let  $N_t$ and $N_r$ denote  the numbers of transmit and receive antennas, respectively.  Time is divided into  baseband sampling intervals, called (time) \emph{slots}. Then  the slot-$t$ realization of the MIMO channel from an active device  to the server can be represented by the $N_r\times N_t$ matrix $\bH_t$.  Given an analog vector-symbol $\bg_t$ transmitted by the active device, the received signal at the server is 
\begin{equation}\label{tx_model_Grass}
\by_t = \sqrt{P} \bH_t \bg_t+ \bw_t
\end{equation}
where $P$ is the transmission power and $\bw_t$ the \emph{additive-white-Gaussian-noise} (AWGN) vector.  In this work, we focus on   transmission of data samples that dominates the data acquisition process.  Their labels have finite values and naturally can be transmitted using digital non-coherent MIMO modulation over a low-rate channel, called \emph{label channel},  orthogonal to the high-rate data channel. Due to its low rate, the label channel can be reasonably assumed to be noiseless similarly as  the CSI feedback channel~\cite{love2008overview}.

\vspace{-5mm}
\subsection{Simulation Models} \label{SimulationModel}
\vspace{-2mm}
Simulation for evaluating learning performance  is based on the following data and channel models. 
The data at different edge devices  are assumed to be \emph{independent and identically distributed} (i.i.d) 
based on the classic  \emph{mixture of Gaussian} (MoG) model, which is  widely adopted in the machine-learning literature. Each data sample is a $1$-by-$L$ complex  random vector. Let $M$ denote the number of data classes. Then the $i$-th sample, denoted as $\bs^{(i)}$, from the $m$-th class  can be modelled as 
\begin{align}
\bs^{(i)} =  \boldsymbol\mu_m + \bz^{(i)}, \qquad \forall i,
\end{align}
where $\boldsymbol\mu_m$ is the mean of the $m$-th class and  $\bz^{(i)} \in {\mathbb{C}}^{1 \times L}$ a deviation vector comprising i.i.d. ${\cal{CN}}(0,\sigma^{2}_{\sf s})$ elements.

Next, high mobility induces temporally correlated MIMO channels. Assuming rich scattering, the classic Clark's model  is applied that translates a speed  into the level of channel temporal correlation. Specifically, within the duration of transmitting a data sample, two realizations of the $(m, n)$-th coefficient of the channel $\bH_t$ separated by $\tau$ slots are correlated with the correlation function given as 
\begin{equation}
\mathbb{E}[(h^{(m,n)}_{t})^{*}h^{(m,n)}_{t + \tau} ] = {\mathcal{J}}_{0}(2\pi f_D\tau),
\end{equation}
where $f_D = \frac{f_c v}{c}$ with $v$ being the speed, $f_c$ carrier frequency and $c$ speed of light, and ${\mathcal{J}}_{0}$ is the zero-th order Bessel function of the first kind. 

\vspace{-3mm}
\section{Fast Analog Transmission Scheme}\label{MainSection}
\vspace{-2mm}
In this section, we discuss two key algorithms  in the proposed FAT scheme, namely GAE and blind analog transmission and detection (see Fig.~\ref{Fig:Grassmann_encoding_scheme}). The received Grassmannian dataset is used for training a classifier model using an existing  Grassmannian classifiaction algorithm such as  \emph{sample Karcher mean} \cite{du2018automatic}, which is adopted in simulation. The details are omitted for brevity.

\vspace{-5mm}
\subsection{Grassmann Analog  Encoding}
\begin{figure}[t]
  \centering
\includegraphics[width=0.55\textwidth]{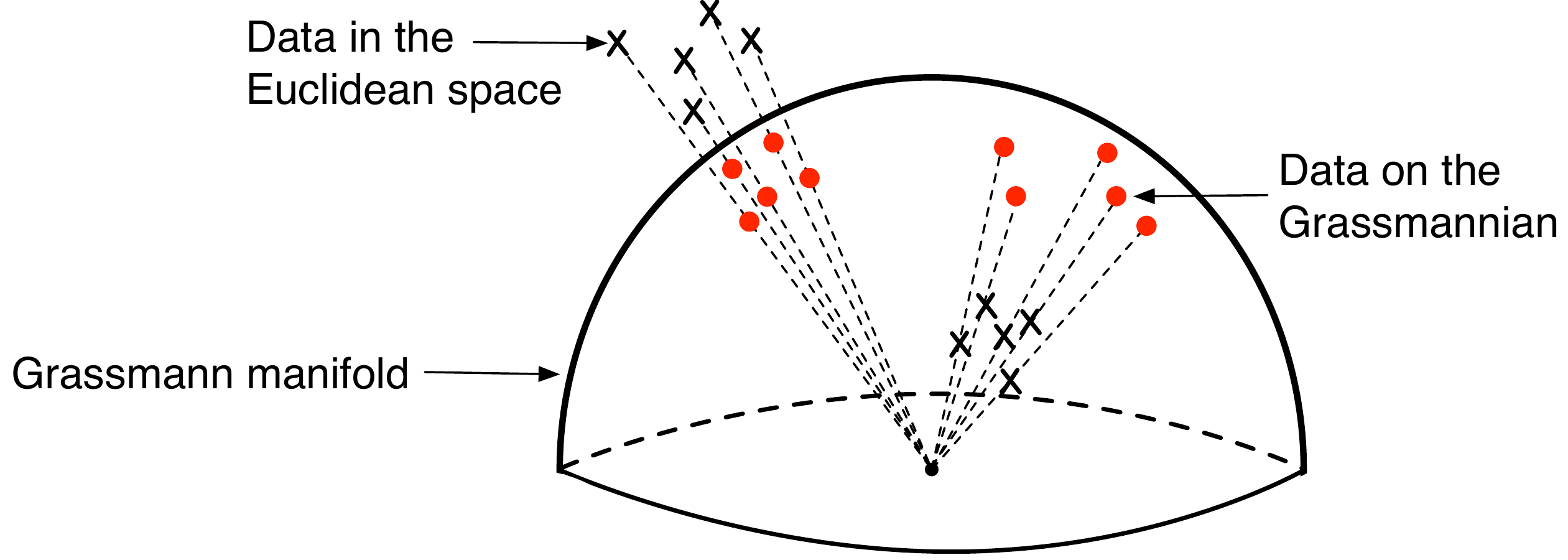} 
     \caption{Principle of Grassmann analog encoding.  }
  \vspace{1mm}
  \label{Fig:GAE}
\end{figure}

\vspace{-2mm}
To facilitate exposition, some mathematical notions are defined   as follows. The $(n, m)$  Grassmann manifold is a set of all $m$-dimensional subspaces in $\mathbb{C}^{n}$, denoted by ${\cal{G}}_{n,m}$~\cite{edelman1998geometry}. For the special case of ${\cal{G}}_{3,1}$, each point on the manifold geometrically corresponds to a unique line passing through the origin as illustrated in Fig.~\ref{Fig:GAE}. 
For ease of notation, a point on ${\cal{G}}_{n,m}$ that is a subspace  is usually represented by an arbitrary basis matrix   spanning the subspace, denoted as $\boldsymbol\Upsilon$. The   subspace distance between two points $\boldsymbol\Upsilon$ and $\boldsymbol\Upsilon'$  on the Grassmannian ${\cal{G}}_{n,m}$, denoted as $d_p(\boldsymbol\Upsilon, \boldsymbol\Upsilon')$, is measured using the commonly used  metric of  \emph{Procrustes distance} for its better performance in simulation: 
\begin{align} \label{eq: procrustes_dist}
d^2_p(\boldsymbol\Upsilon, \boldsymbol\Upsilon') = m- \text{tr}\l\{ \boldsymbol\Upsilon\boldsymbol\Upsilon^{H}\boldsymbol\Upsilon'(\boldsymbol\Upsilon')^{H}\r\}.
\end{align}

As discussed, GAE at the active device  endows on  FAT the channel-invariant property, thereby enabling blind analog transmission with robustness against fast fading.  As illustrated in Fig.~\ref{Fig:GAE}, the mathematical  principle of GAE is to project original data samples (vectors in the Euclidean space) onto  the Grassmann manifold, generating subspace matrices as the output.   The GAE algorithm  is described  as follows. 

{\bf Step 1 (Vector-to-matrix conversion)}: Consider a  data sample  that is a $1\times L$   row vector, say  $\bs^{(i)}$, with $L$ being an  integer multiple of the number of transmit antennas $N_t$. Then $\bs^{(i)}$ can be divided into $1\times T $ sub-vectors with $T = L/N_t$: $\bs^{(i)} = [\bs^{(i)}_1, \bs^{(i)}_2, \cdots, \bs^{(i)}_{N_t}]$.  It follows that  $\bs^{(i)}$ can be converted into  a   $N_t\times T$ data matrix ${\bX}^{(i)} $ having $\{\bs^{(i)}_n\}$ as rows. The matrix ${\bX}^{(i)} $ such constructed is typically \emph{fat} ($N_t < T$) since a data-sample vector is usually long ($L \gg N_t$).  For the case where $L$ is not an integer multiple of $N_t$, zero-padding can be applied to lengthen $\bs^{(i)}$ so that the integer-multiple constraint is met.

{\bf Step 2 (Projection onto Grassmannian)}: The key step in encoding is to project the matrix ${\bX}^{(i)} $ constructed in the preceding step onto a single point on the Grassmannian ${\cal{G}}_{T, N_t}$. To this end, decompose the matrix ${\bX}^{(i)} $ by  \emph{singular value decomposition} (SVD) as ${\bX}^{(i)}  = {\bU}^{(i)} \boldsymbol{\Sigma}^{(i)} {\bG}^{(i)}$. Then ${\bG}^{(i)}$ is a $N_t \times T$ basis matrix spanning the row space of ${\bX}^{(i)}$ and thus a point on ${\cal{G}}_{T, N_t}$. The encoder uses  ${\bG}^{(i)}$ as the output from encoding the data-sample $\bs^{(i)}$. 

In summary, the advantages of GAE are threefold: 1) enabling  blind transmission and detection as discussed in the next sub-section, 2) endowing on edge learning robustness against data distortion by fast fading as shown in simulation results, and 3) allowing seamless integration with learning on Grassmannian without signal decoding and conversion.  

\vspace{-5mm}
\subsection{Blind Analog Transmission and Detection}
\vspace{-2mm}
Given Grassmann encoding, the procedures for subsequent blind analog  transmission and detection in FAT  are described as follows. 

{\bf Step 1 (Analog transmission)}:  After encoding each data-sample, say $\bs^{(i)}$, into a subspace basis matrix ${\bG}^{(i)}$, the $N_t\times T$ matrix ${\bG}^{(i)}$ is directly transmitted by the active device  over $T$ slots using linear analog modulation and the array of $N_t$ antennas. To reflect channel temporal variation, it is necessary write ${\bG}^{(i)}$ in terms of its columns: ${\bG}^{(i)} = [{\bg}^{(i)}_1, {\bg}^{(i)}_2, \cdots, {\bg}^{(i)}_T]$. Then the  received signal due to the transmission of ${\bG}^{(i)}$ can be represented by the $N_r\times T$ matrix $\bY^{(i)} = [\by^{(i)}_1, \by^{(i)}_2, \cdots, \by^{(i)}_T]$ with $\by^{(i)}_t$ given as 
\begin{equation}\label{tx_model_Grass}
\by^{(i)}_t  = \sqrt{P} \bH^{(i)}_t \bg^{(i)}_t + \bw^{(i)}_t. 
\end{equation}
For continuous time-shared distributed  uploading of total $N$ data samples, $t = 1, 2, \cdots N T$. It is important to observe from \eqref{tx_model_Grass} that due to high mobility, the channel  $\{\bH^{(i)}_t\}$ varies in the $T$-slot transmission duration of a single data sample, which has a negative effect on decoding as discussed in the sequel.  

{\bf Step 2 (Grassmann analog detection)}: The detection of the transmitted  encoded analog   space-time symbol ${\bG}^{(i)}$ involves the extraction of the row space, denoted by the $N_t \times T$ unitary matrix $\widehat{\bG}^{(i)}$, from the SVD of the received $N_r\times T$ space-time signal $\bY^{(i)}$ specified in \eqref{tx_model_Grass}, namely $\bY^{(i)} = \bV^{(i)}\boldsymbol{\Pi}^{(i)}    \widehat{\bG}^{(i)}$. Consider the special case of zero noise and static channel. The detected symbol $\widehat{\bG}^{(i)} =  \mathbf{O} \bG^{(i)}$ where $\mathbf{O}$ a $N_t\times N_t$ rotation (unitary) matrix. In other words, $\widehat{\bG}^{(i)}$ and $\bG^{(i)}$ are the identical point on the Grassmannian, corresponding to perfect detection. In the presence of noise and channel variation, they are two different points and the resultant detection error affects learning. Based on the above detection procedure, the output training dataset, called \emph{Grassmann dataset},  is a sequence of $N$ labeled subspace matrices (points on the Grassmannian),  
$[\widehat{\bG}^{(1)}, \widehat{\bG}^{(2)}, \cdots, \widehat{\bG}^{(N)}]$, whose labels are acquired by the server via the said low-rate label channel. 

\begin{remark}[Blind Transmission and Detection] \emph{Both the analog transmission and detection in the above steps are independent of the channel. In particular, the detection of Grassmann dataset involves SVDs of the received array observations that do not require any channel knowledge.} 
\end{remark}

\vspace{-5mm}
\section{Understanding the Design}\label{Sec:Understanding_Design}
\vspace{-2mm}
\subsection{Grassmann Analog Encoding  Preserves Clustering} An important reason FAT supports edge classification is that \emph{GAE retains the class structure in the original dataset}. This property  is illustrated in Fig.~\ref{Fig:datastructure} where the high-dimensional datasets are visualized in the 2D plane using a well known visualization algorithm, \emph{t-distributed stochastic neighbour embedding} (t-SNE). 
As discussed in the sequel, GAE incurs  DoF loss in the dataset. Consequently, one can observe form Fig.~\ref{Fig:datastructure} that data classes are less compact after GAE, sacrificing some level of discriminant of the dataset. The loss,   nevertheless,  yields  communication advantages discussed shortly. 

\vspace{-5mm}
\subsection{Trading DoF Loss for Robustness and Low Latency}
\vspace{-2mm}
The Grassmann encoding design leads to the DoF loss (or discriminant  loss), which may make data points among different classes that are well-separated in the Euclidean space become much closer or even overlapped on the Grassmannian. The loss has a negative effect on  the classification performance. The phenomenon is illustrated by the following  example. Besides SVD, an alternative method for GAE is  LQ decomposition. Consider the LQ decomposition of two data matrices $\bX = \bL_{\bX}\bG_{\bX}$ and $\bY = \bL_{\bY}\bG_{\bY}$, where the unitary matrices  $\bG_{\bX}$ and $\bG_{\bY}$ represent identical subspaces (or identical encoding outputs) as the SVD counterparts and  $\bL_{\bX}$ and $ \bL_{\bY}$ are lower triangular matrices. If $\text{span}(\bG_{\bX}) = \text{span}(\bG_{\bY})$ but $\bL_{\bX}\neq \bL_{\bY}$, the Euclidean distance between $\bX$ and $\bY$ is $d^2_E(\bX,\bY) \neq   0$. 
However, based on  \eqref{eq: procrustes_dist}, the Procrustes distance  between the GAE encoded data samples, namely $\bG_\bX$ and $\bG_\bY$, is $d^2_p(\bG_{\bX},\bG_{\bY}) = 0$. 

A key finding in this work is that the DoF loss of GAE is more than compensated by its robustness against fast fading that can cause severe errors in data transmission without GAE. As a  result, GAE leads to a net performance gain over conventional schemes at high mobility. Furthermore, GAE also leads to transmission-latency reduction as it eliminates  channel-training overhead  and enables analog transmission faster than digital counterparts.

%\begin{figure}[tt]
%  \centering
%\subfigure[Data structure before GAE]{\includegraphics[width=0.29\textwidth]{tsne_coherent.pdf} \label{Sensing data structure}}
%\subfigure[Data structure after GAE]{\includegraphics[width=0.29\textwidth]{tsne_Grassmann.pdf}\label{Grassmannian data structure} }
%     \caption{The  clustering  structure of a binary MoG data set   (a) before and (b) after GAE.}
% % \vspace{-1mm}
%  \label{Fig:datastructure}
%\end{figure}

%\begin{figure}[tt]
%  \centering
%\includegraphics[width=0.6\textwidth]{tsne.pdf} 
%     \caption{The  clustering  structure of a binary MoG data set   (a) before and (b) after GAE.}
% % \vspace{-1mm}
%  \label{Fig:datastructure}
%\end{figure}

\begin{figure}[t]
\centering
\begin{minipage}{0.495\textwidth}
\centering
\includegraphics[width=1.1\textwidth]{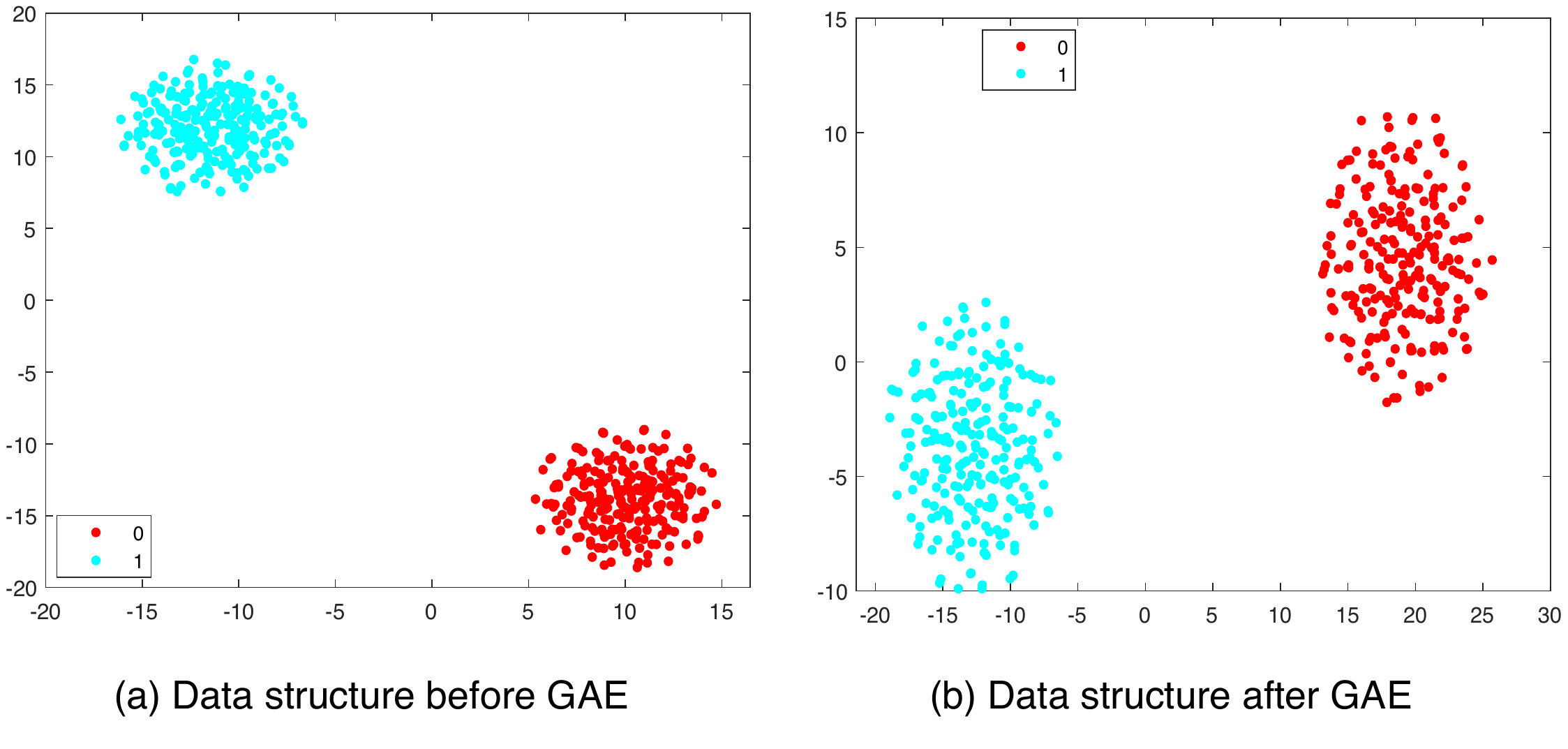}
\caption{The  clustering  structure of a binary MoG data set   (a) before and (b) after GAE. }
  \label{Fig:datastructure}
\end{minipage}
\begin{minipage}{0.495\textwidth}
\centering
\includegraphics[width=0.75\textwidth]{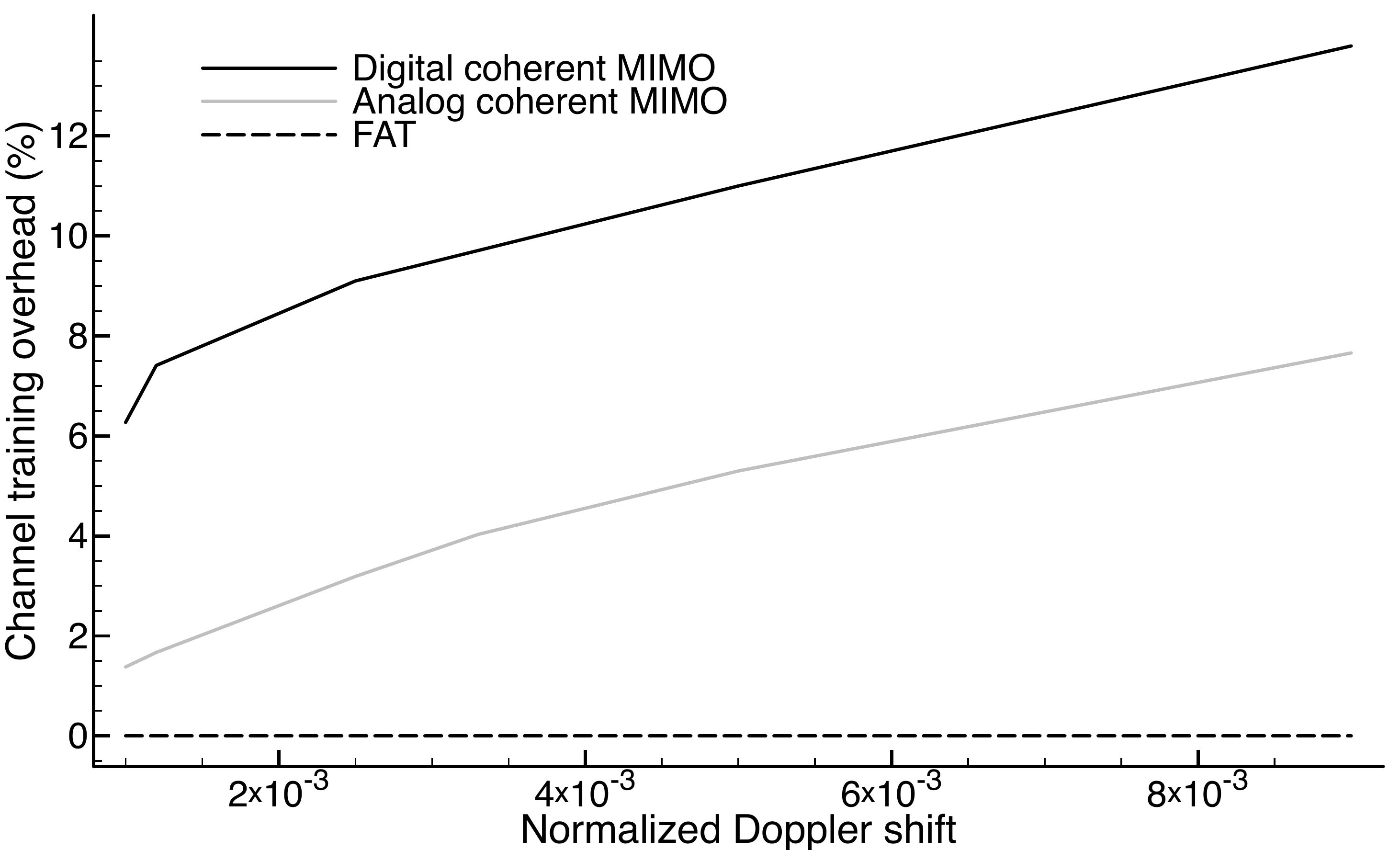}
\caption{The channel-training overhead versus  normalized Doppler shift for the  target  classification error rate of $1\times 10^{-3}$.}
 \label{Fig:tradeoff}
\end{minipage}
%\vspace{-30pt}
\end{figure}

\vspace{-5mm}
\section{Simulation results}\label{Simulation}
\vspace{-2mm}
The simulation parameters are set as follows unless specified otherwise. The number of Gaussian classes is $M =  2$ with source data parametric ratio, i.e. $\parallel\boldsymbol\mu_{m}\parallel^2/\sigma^2_{\sf s}$, being $15$ dB and the dimension of each data sample  is $L = 48$. The $4 \times 2$ MIMO channel is temporally correlated with the variation speed  specified by the normalized Doppler shift  $f_DT_{s} = 0.01$ with $T_{s}$ being the baseband sampling interval (or time slot). Define the training and test datasets are generated based on the discussed MoG model, which  comprise  $200$ and $2000$ samples, respectively.

The performance of FAT is benchmarked against  two high-rate coherent schemes: digital  and analog MIMO transmission, both of which assume a  MMSE linear receiver and thus require channel training to acquire the needed CSI. Like FAT, analog MIMO transmits data samples directly by linear analog modulation. On the other hand, digital MIMO quantizes data samples into $8$-bit per coefficient and modulates each symbol  using QPSK before MIMO transmission. All considered schemes have no error control coding. 
%Their performance can be  improved by such coding e.g., convolutional coding and Shannon-Kotel'nikov mapping for digital and analog transmission, respectively, which is outside the scope of this paper.

%\begin{figure}[tt]
%  \centering
%\subfigure[Effect of training dataset size]{\includegraphics[width=0.42\textwidth]{SampleSize_Accuracy_v2.pdf} \label{DatasetSize}}
%\subfigure[Effect of DoF loss ]{\includegraphics[width=0.42\textwidth]{effect_DoF_loss_v3.pdf}\label{Fig:effect_DoF_loss} }
%     \caption{The dependence of classification error rate on (a) training dataset size and (b) DoF loss due to GAE.}
% % \vspace{-1mm}
%%  \label{Fig:SampleSize}
%\end{figure}

%\subsection{Effect of the DoF loss}
%In Fig.~\ref{Fig:effect_DoF_loss}, the DoF loss effect of the proposed GAE on the classification performance is evaluated, where the DoF loss percentage is defined as $r = \frac{(N_t)^2}{D}\times 100\%$. One can observe that as the DoF loss percentage increases, the classification performance on the Grassmannian degrades accordingly. This coincides with our previous analysis that non-overlapping points among different classes in the Euclidean space may become much closer or even overlapped on the Grassmannian. 

\vspace{-5mm}
\subsection{Communication Latency Performance}
\vspace{-2mm}
While FAT is free of  channel-training,  benchmark schemes incur training overhead that can be quantified by  the fraction of a frame allocated for the purpose  i.e., the ratio $P/(P+D)$ with $P$ and $D$ illustrated in Fig.~\ref{Fig:latency}. The curves of overhead versus Doppler shift are displayed in Fig.~\ref{Fig:tradeoff} for FAT and two mentioned benchmarking schemes for a given classification-error rate of $1\times 10^{-3}$. One can observe that the overhead grows monotonically with the Doppler shift as the channel fading becomes faster. For high-mobility with Doppler approaching  $10^{-2}$ , the overhead can be more than $12\%$ and $6\%$ for digital and analog coherent MIMO, respectively. Furthermore, given the same performance, digital coherent MIMO (with QPSK modulation and $8$-bit quantization) requires $4$ times more frames for transmitting the training dataset than the two analog schemes. This suggests that analog transmission is preferable  for data acquisition targeting  edge learning.

%\begin{figure}[tt]
%  \centering
%\includegraphics[width=0.4\textwidth]{Latency_vs_Doppler_v4.pdf} 
%     \caption{The channel-training overhead versus  normalized Doppler shift for the  target  classification error rate of $1\times 10^{-3}$.}
% % \vspace{-1mm}
%  \label{Fig:tradeoff}
%\end{figure}

\vspace{-5mm}
\subsection{Learning Performance}
\begin{figure}[tt]
\centering
\subfigure[Effect of Doppler shift ]{\label{Fig:doppler_shift}\includegraphics[width=0.4\textwidth]{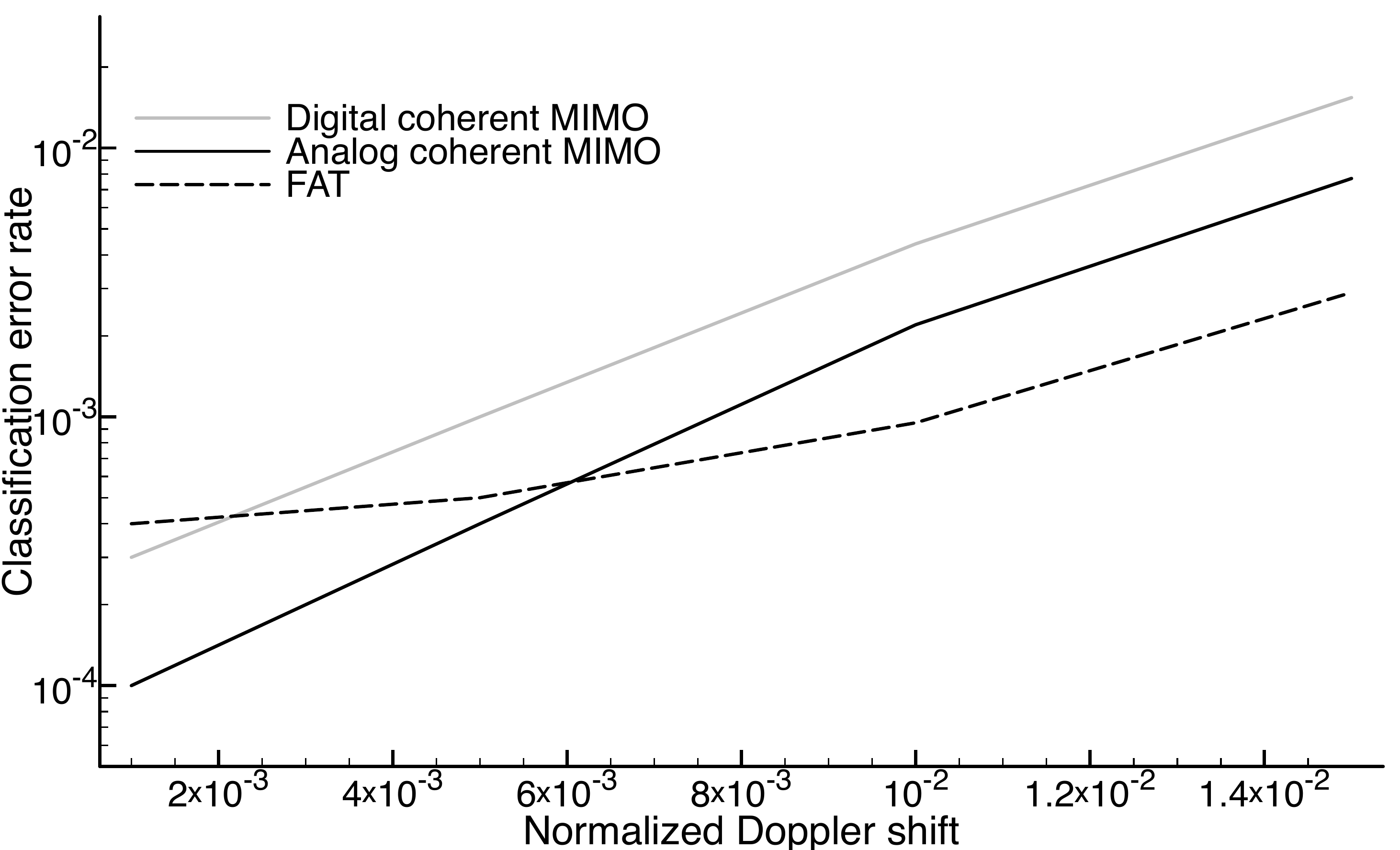}}
\subfigure[Effect of average transmit SNR]{\label{Fig:effect_channel_noise}\includegraphics[width=0.4\textwidth]{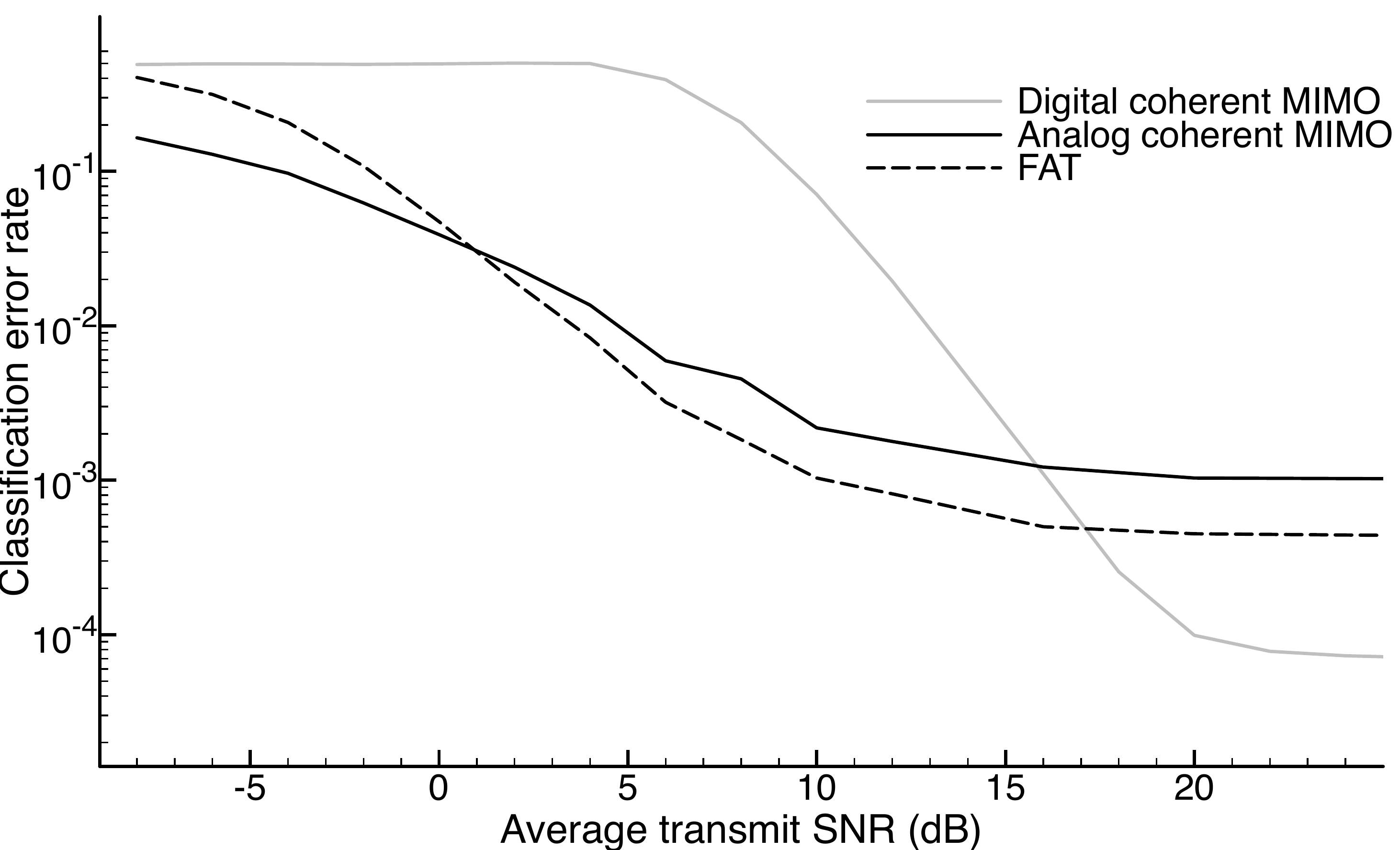}}
\caption{Learning performance comparison for two cases: (a) a varying Doppler shift with the average transmit SNR equal to $15$ dB;  (b) a varying average transmit SNR with the normalized Doppler shift fixed at $0.01$. 
 }
\label{Fig:Non-coherent v.s. coherent transmission}
\vspace{2mm}
\end{figure}

\vspace{-2mm}
Classifier models discussed in Section~\ref{MainSection} are trained  using the training dataset acquired using  different transmission schemes and then evaluated using the test dataset. The resultant classification error rates are compared in Fig.~\ref{Fig:Non-coherent v.s. coherent transmission}  by  varying Doppler shift and average transmit SNR. Several observations can be made. In the range of moderate to large Doppler shift (i.e., larger than $6\times 10^{-3}$), the proposed FAT outperforms the benchmarking schemes, supporting the former's intended application in high-mobility  data acquisition. Furthermore, at high mobility (i.e., Doppler equal to $0.01$), FAT achieves the best performance  in  the practical SNR range ($0 - 15$ dB). On the other hand, analog and digital coherent MIMO are preferred at low and high SNRs, respectively. The above observations reconfirm the conclusion from preceding latency comparison that analog transmission (especially FAT) is a  promising solution  for high-mobility data acquisition for edge learning. 

%\section{Concluding Remarks}
%In this paper, we have presented the scheme of \emph{fast analog transmission} (FAT) that enables high-mobility wireless data acquisition for edge learning. The scheme features the novel technique of Grassmannian analog encoding. Though we considers in this work  a relatively simple learning task of data clustering, FAT can support general   AI models that is capable of processing  Grassmannian data such as deep neural networks on the Grassmannian manifold, which belongs to the increasingly popular area of geometry aware deep learning.  

\vspace{-5mm}
\bibliography{reference}
\vspace{-2mm}
\bibliographystyle{IEEEtran}

\end{document}